\documentclass[referee]{aa} 
\usepackage{graphicx}
\usepackage{natbib}
\usepackage{txfonts}
\usepackage{ulem} 
\begin{document}
   \title{Cold gas as an ice diagnostic toward low mass protostars }

   \author{Karin I. \"Oberg\inst{1}
          \and
		  Sandrine Bottinelli\inst{1}
		  \and
          Ewine F. van Dishoeck\inst{1,2}
          }

   \institute{Leiden Observatory, Leiden University, P.O. Box 9513, NL 2300 RA Leiden, The Netherlands
         \and
             Max-Planck-Institut f\"ur extraterrestrische Physik (MPE), Giessenbachstraat 1, 85748 Garching, Germany\\
             }

   \date{}

 
  \abstract
{Up to 90\% of the chemical reactions during star formation occurs on ice surfaces, probably including the formation of complex organics. Only the most abundant ice species are however observed directly by infrared spectroscopy. }
{This study aims to develop an indirect observational method of ices based on non-thermal ice desorption in the colder part of protostellar envelopes. }
{The IRAM 30m telescope was employed to observe two molecules that can be detected both in the gas and the ice, CH$_3$OH and HNCO, toward 4 low mass embedded protostars. Their respective gas-phase column densities are  determined using rotational diagrams. The relationship between ice and gas phase abundances is subsequently determined.}
{The observed gas and ice abundances span several orders of magnitude. Most of the CH$_3$OH and HNCO gas along the lines of sight is inferred to be quiescent from the measured line widths and the derived excitation temperatures, and hence not affected by thermal desorption close to the protostar or in outflow shocks. The measured gas to ice ratio of $\sim$10$^{-4}$ agrees well with model predictions for non-thermal desorption under cold envelope conditions and there is a tentative correlation between ice and gas phase abundances. This indicates that non-thermal desorption products can serve as a signature of the ice composition. A larger sample is however necessary to provide a conclusive proof of concept.}
{}


   \maketitle
%

\section{Introduction}

In cold pre-stellar cores, more than 90\% of all molecules, except for H$_2$, are found in ices \citep{Caselli99, Bergin02}. These ices build up through accretion of atoms and molecules onto cold (sub)micron-sized silicate particles and subsequent hydrogenation to form e.g. H$_2$O from O \citep{Leger85,Boogert04}. Observations show that H$_2$O is the main ice constituent in most lines of sight, with a typical abundance of $1\times 10^{-4}$ with respect to H$_2$, followed by CO, CO$_2$ and CH$_3$OH \citep{Gibb04, Pontoppidan04}.

During star formation, these ices may be modified by interactions with cosmic rays, UV irradiation, and heating to form complex organic species \citep{Garrod08}. Gas phase complex species have been observed toward several high and low mass protostars, so-called hot cores and corinos \citep{Bisschop07, Bottinelli07}. Whether these molecules are formed in the ice and subsequently evaporated, or formed in the hot gas phase from desorbed simpler ices such as CH$_3$OH is still debated. This is not easily resolved because the abundances of the solid complex molecules are too low to be detected with infrared observations of ices even if they are present in the ice. Therefore, observing gas-phase abundances in the cold envelope may be the most robust constraint on complex ice processes available.

Experimental investigations have concluded that non-thermal desorption is efficient for several common ice molecules, such as CO, CO$_2$, and H$_2$O, with photodesorption yields of $\sim$10$^{-3}$ per incident photon \citep{Westley95, Oberg08b, Oberg09}. Photodesorption is possible inside cold dark cloud cores and protostellar envelopes because of constant UV fields generated from cosmic ray interactions with H$_2$ \citep{Shen04}. Thus a small, but significant, part of the molecules formed in the ice should always be present in the gas phase. This explains observed abundances of gas phase CH$_3$OH in translucent clouds, dark cloud cores and protostellar envelopes \citep{Turner98, Maret05,Requenatorres07}. The amount of CH$_3$OH gas observed in these environments suggests that complex molecules (e.g. methyl formate) that form in the ice should be observable in the gas phase due to ice photodesorption, if their abundance ratios with respect to CH$_3$OH in the ice are the same as observed in hot cores and corinos.

For the first time, we combine infrared ice observations and millimeter gas observations for the same lines of sight to investigate the connection between ice and quiescent gas abundances. We focus on the only commonly observed ice components that have rotational transitions in the millimeter spectral range -- CH$_3$OH and HNCO. The CH$_3$OH ice abundances in low mass protostellar envelopes vary between 1--30\% with respect to H$_2$O ice \citep{Boogert08}. It is also one of the most common hot corino gas phase molecules with typical abundances of $10^{-7}-10^{-6}$ with respect to H$_2$. HNCO gas is also commonly detected in hot cores. Solid HNCO (in the form of OCN$^-$) is only detected toward a few low mass protostars, but strict upper limits exist for more, resulting in an abundance span of an order of magnitude \citep{vanBroekhuizen05}. These large variations in ice abundances imply that CH$_3$OH and HNCO are appealing test cases for our theory that quiescent complex gas abundances reflect the composition of the co-existing ice mantles, under the assumption that OCN$^-$ is protonated during desorption. 

We have observed gas phase CH$_3$OH and HNCO with the IRAM 30m toward four low mass protostars for which CH$_3$OH and HNCO ice detections or upper limits already exist. Two of these sources also have OCN$^-$ ice upper limits. These sources are complemented with literature values to constrain further the relationship between ice and gas phase abundances.

\section{Source selection}

\begin{table*}[htp]
\begin{center}
\caption{Targets with pointing positions and ice data.}             
\label{sourcedata}      
\centering                          
\begin{tabular}{l c c c c c c c c}        
\hline\hline                 
Source &RA&Dec&$V_{\rm LSR}$&$\alpha$&cloud&H$_2$O col. dens. &CH$_3$OH&HNCO \\    
&&&km s$^{-1}$&2--24 $\mu$m&&$10^{18}$cm$^{-2}$&\% H$_2$O&\% H$_2$O\\
\hline      
IRAS 03254+3050&03:26:37.45&+30:51:27.9&5.1&0.90&Perseus&3.66&$<$4.6&--\\
B1-b&03:33:20.34&+31:07:21.4&6.5&0.68&Perseus&17.67&11.2&--\\
L1489 IRS&04:04:43.37&+26:18:56.4&7.1&1.10&Taurus&4.26&4.9&$<$0.06\\
SVS 4-5&18:29:57.59&01:13:00.6&7.8&1.26&Serpens&5.65&25.2&$<$0.27\\
\hline                                   
\end{tabular}
\end{center}
\end{table*}

The four sources IRAS~03254+3050, B1-b, L1489~IRS and SVS~4-5 were chosen from the `cores to disks' ($c2d$) sample of low mass protostars with ice detections \citep{Boogert08} to span CH$_3$OH abundances of $4-25$\% with respect to H$_2$O. The $c2d$ sample partly overlaps with an earlier ground based survey using the VLT, for which the OCN$^-$ abundances and upper limits were determined \citep{vanBroekhuizen05} and two of the sources have OCN$^-$ ice upper limits (Table~\ref{sourcedata}).

According to the classification scheme of Lada \& Wilking (1984), all sources are embedded class 0/I sources with spectral energy distribution (SED) slopes in the mid-infrared 2--24 $\mu$m between 0.68 and 1.26. Their envelopes are of similar mass, as traced by the H$_2$O ice abundance, with the possible exception of B1-b, which has a factor of 3 higher column density. Except for SVS~4-5 the sources are isolated on the scale of the IRAM 30m beam. Both B1-b and L1489 IRS have however moderate outflows associated with them that may contribute to the detected lines \citep{Jorgensen06,Girart02}.

SVS~4 region is one of the densest nearby star-forming regions and SVS~4-5 is located $\sim$20'' away from the class 0 low-mass protostar SMM~4, which has a large envelope and an associated outflow. In addition, the young stellar objects SVS~4-2--12 are all located within 30$''$ of SVS~4-5 and the emission from SVS~4-5 is probably contaminated by emission from its surroundings when observed with a beam of width as large as 24''  (as is the case here). Despite these complications in interpreting the data, SVS~4-5 is included in the sample because of its unusually high CH$_3$OH ice abundance, which was determined from ice mapping of the SVS~4 region by \citet{Pontoppidan04}. 

\section{Observations}

\begin{table*}
\begin{center}
\caption{Observed frequencies and targeted molecules at the four different settings.}             
\label{transitions}      
\centering                          
\begin{tabular}{l c c c c c c c c c c c }        
\hline\hline                 
Frequency&Frequency&Transition$^{\rm a}$&$E_{\rm u}$ (K)&rms (mK) &\multicolumn{4}{c} {Integrated intensity [uncertainty] (K km s$^{-1}$)}\\    
range (GHz)&(GHz)&&&&IRAS 03254&B1-b&L1489 IRS&SVS 4-5\\
\hline      
CH$_3$OH\\
1) 96.739--96.756&96.739 (E$^-$)&2$_{1 2 }$ -- 1$_{ 1 1 }$ &20.0&11--15&0.11 [0.03]&1.65 [0.22]&0.10 [0.03]&1.47  [0.33]\\
&96.741 (A$^+$)&2$_{ 0 2 }$-- 1$_{ 0 1 }$ &14.4&&0.16 [0.04]&2.18 [0.21]&0.17 [0.05]&2.72 [0.58]\\
&96.745 (E$^+$)&2$_{ 0 2 }$-- 1$_{ 0 1 }$ &27.5&&0.017 [0.017]&0.37 [0.20]&0.024 [0.024]&0.64 [0.39]\\
&96.756 (E$^+$)&2$_{ 1 1 }$ -- 1$_{ 1 0 }$ &35.4&&$<$0.01&0.09 [0.26]&$<$0.024 &0.074 [0.030]\\
2) 251.360--251.811&251.738 (A$^{\rm \pm}$)&6$_{ 3 3 }$ -- 6$_{ 2 4 }$ &98.6&28--40&$<$0.033&--&$<$0.079&$<$0.26\\
\hline
HNCO\\
3) 109.872--109.938&109.906&5$_{ 0 5 }$ -- 4$_{ 0 4 }$&15.4&15--23&0.038 [0.030]&--&$<$0.027&0.22 [0.07]\\
4) 131.885--131.886&131.886&6$_{ 0 6 }$ -- 5$_{ 0 5 }$&18.6&19--29&$<$0.021&0.43 [0.06]&$<$0.037&0.22 [0.07]\\ 
\hline                                   
\end{tabular}
\end{center}
$^{\rm a}$ The quantum numbers for the pure rotational transitions of CH$_3$OH and HNCO are $J_{\rm K_aK_c}$ and $J_{\rm K_{-1}K_{+1}}$, respectively.
\end{table*}

The observations were carried out in March 2008 with the 30-m telescope of the Institut de RadioAstronomie Millim\'etrique (IRAM). The positions used for pointing are listed in Table~\ref{sourcedata}. The line frequencies are taken from the JPL molecular database. Although the observations were centered on the protostars themselves, the choice of low excitation lines and relatively large beams ensures that the cold outer envelope is almost completely sampled. Specifically, we targeted CH$_3$OH transitions with $E_{\rm up}$, the energy of the upper level of the transition, between 7 and 100~K. The HNCO lines were observed after the preliminary reduction of the CH$_3$OH data, and because of its low excitation temperature we chose to observe two of the lowest lying HNCO transitions with $E_{\rm up}$ of 15 and 19~K. The observations were carried out using four different receiver settings with the frequency ranges shown in Table~\ref{transitions}. Each receiver was  connected to a unit of the autocorrelator, with spectral resolutions of 80 or 320 kHz and bandwidths between 80 and 480 MHz, equivalent to a velocity resolution of 0.3, 0.4  and 0.2 km s$^{-1}$ in settings 1, 2, and 3/4, respectively. Typical system temperatures were 100-200 K, 200-500 K, and 700-1000 K, at 3, 2, and 1 mm, respectively.

All observations were carried out using wobbler switching with a 110$''$ throw in azimuth. Pointing and focus were regularly checked using planets or strong quasars, providing a pointing accuracy of 3$''$. All intensities reported in this paper are expressed in units of main-beam brightness temperature, which were converted from antenna temperatures using main beam efficiencies of 76, 69, and 50\%, at 3, 2, and 1 mm. At these wavelengths, the beam sizes were 24, 16, and 10$''$, respectively.

\section{Results}

\begin{figure}
\resizebox{\hsize}{!}{\includegraphics{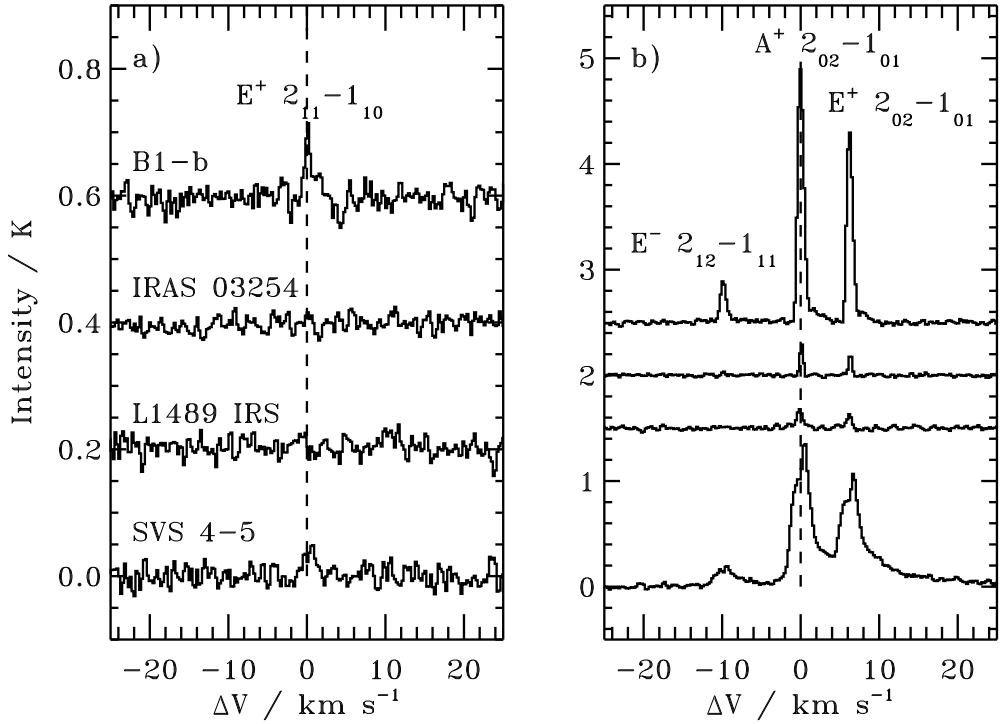}}
\caption{The observed CH$_3$OH lines in setting 1 toward the four low mass protostars plotted versus $\Delta$V, the deviation from the source V$_{\rm lsr}$. The data in the left panel are centered on a rest frequency of 96.756~GHz and in the right panel on a rest frequency of 96.741~GHz.}
\label{sp_ch3oh}
\end{figure}

\begin{figure}
\resizebox{\hsize}{!}{\includegraphics{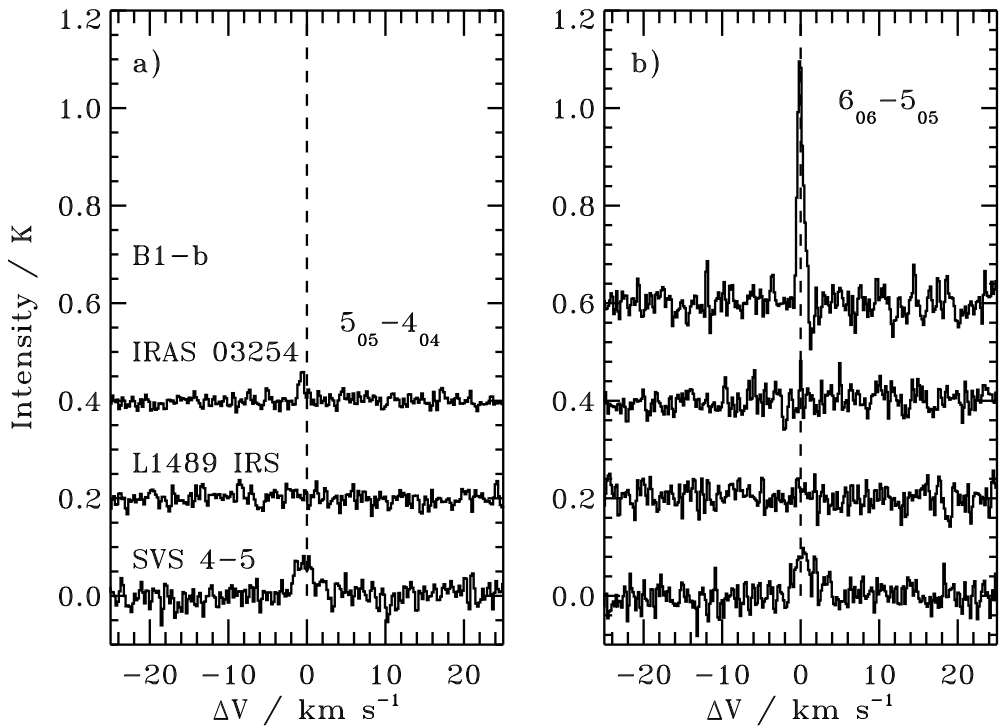}}
\caption{The observed HNCO lines in setting 3 and 4 plotted versus $\Delta$V, the deviation from the source V$_{\rm lsr}$. The data in the left panel are centered on a rest frequency of 109.906 GHz and in the right panel on 131.886~GHz.}
\label{sp_hnco}
\end{figure}

\begin{figure}
\resizebox{\hsize}{!}{\includegraphics{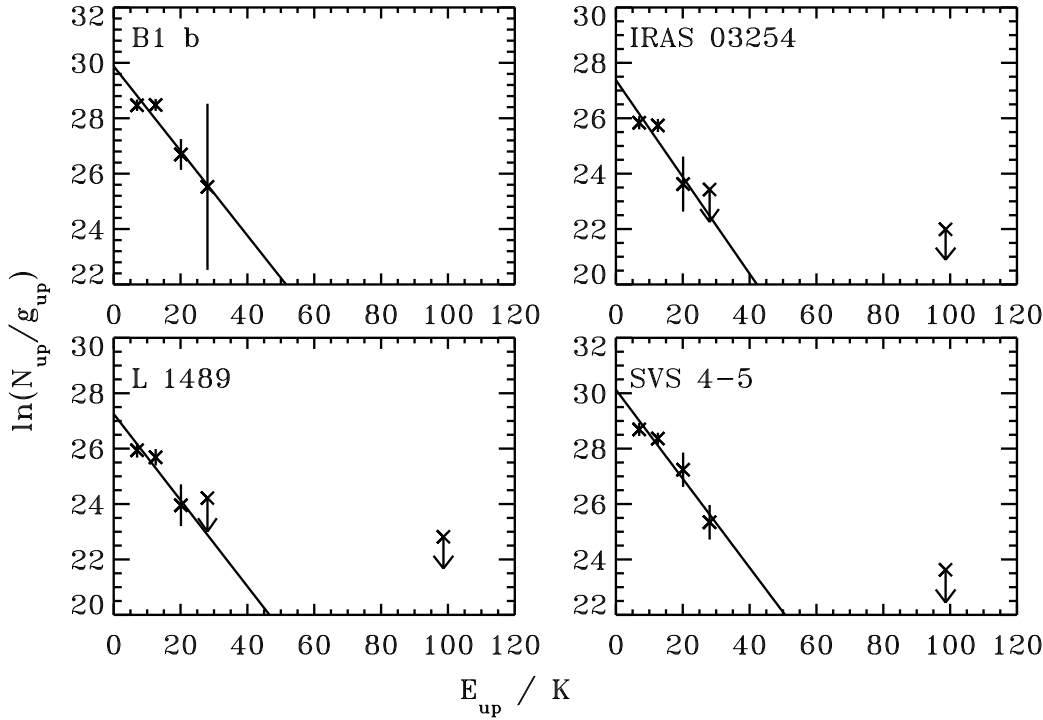}}
\caption{CH$_3$OH rotation diagrams including detections and upper limits. }
\label{rot_ch3oh}
\end{figure}

\begin{figure}
\resizebox{\hsize}{!}{\includegraphics[trim = 0in 2.3in 0in 0in]{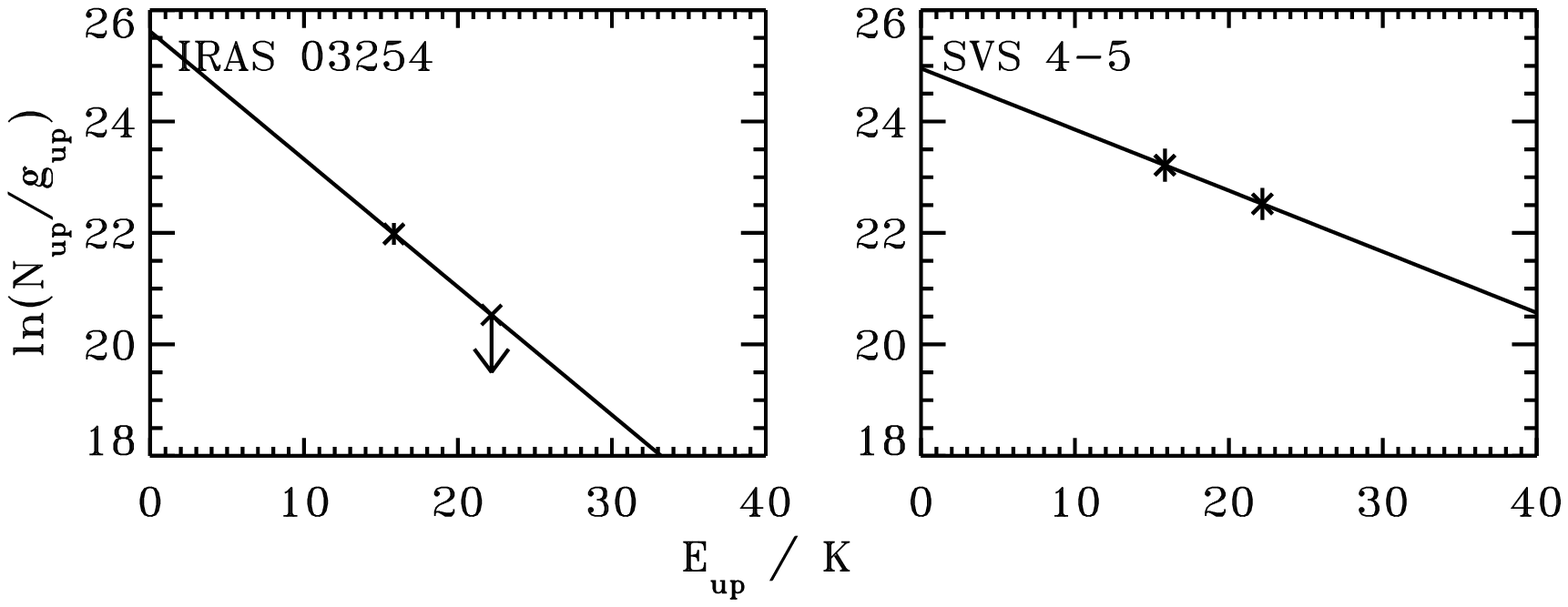}}
\caption{HNCO rotation diagrams where both lines are detected or the upper limit is strict.  }
\label{rot_hnco}
\end{figure}

Figure~\ref{sp_ch3oh} shows the spectra derived for setting 1, which is the only setting in which CH$_3$OH is  detected.  Figure~\ref{sp_hnco} shows the obtained spectra in settings 3 and 4, with all targets observed in both settings except for B1-b. The observed lines were fitted with a single Gaussian to calculate the line widths and integrated intensities in Table \ref{transitions}. The Gaussian fits were restricted to exclude the wings observable for B1--b and SVS~4--5. The resulting line widths range from 0.4 to 4.0 km s$^{-1}$, but three of the sources (B1-b, IRAS 03254, and L1489~IRS) consistently have line widths of below 1 km s$^{-1}$ (Table~\ref{gasdata}). Coupled with the low excitation temperatures (below), we most likely probe the quiescent envelope rather than outflows or hot corinos. The fourth source SVS~4--5 has several emission components, reflecting the complexity of the SVS~4 region and containing contributions from non-quiescent gas in for example the nearby outflow from SMM~4. The typical envelope angular size for the source distance is $\sim$1', which is larger than the largest beam size. Hence, we assume in the analysis that there is no beam dilution. 

In Figs.~\ref{rot_ch3oh} and~\ref{rot_hnco}, we use the rotational diagram method \citep{Goldsmith99} to derive rotational temperatures and column densities. The relations evident in the CH$_3$OH diagrams are approximately linear, with a possible slight deviation for the one CH$_3$OH A detection. A line is fitted to all CH$_3$OH detections, with the assumption that  the populations of E and A species are approximately equal. The 2$\sigma$ upper limits (derived from the rms in Table \ref{transitions}) are overplotted to enable us to ensure that they do not provide further constraints on the fitted line. Except for SVS~4-5, HNCO column densities were derived using the CH$_3$OH rotational temperatures. The lower limit to the CH$_3$OH temperature of 4~K was used for IRAS 03254 to accommodate the strict upper limit from setting 3 (Table \ref{gasdata}). The resulting CH$_3$OH and HNCO temperatures vary between 4 and 9 K and the CH$_3$OH and HNCO column densities vary between $1.8-27\times 10^{13}$ and $0.095-2.4\times 10^{13}$ cm$^{-2}$, respectively. CH$_3$OH is easily sub-thermally excited at the expected densities in outer protostellar envelopes of approximately 10$^4$ cm$^{-3}$ and hence the rotational temperature cannot be directly translated into a kinetic temperature \citep{Bachiller95}.

In previous studies, gas phase CH$_3$OH was observed toward three other low mass protostars that were also observed by Spitzer to study ices \citep{Boogert08}: Elias 29, R CrA 7A, and B (Table \ref{gasdata}). The line widths imply that the observed CH$_3$OH lines toward Elias 29 trace quiescent material, while those observed toward R CrA 7A and B do not.

\begin{figure}
\resizebox{\hsize}{!}{\includegraphics{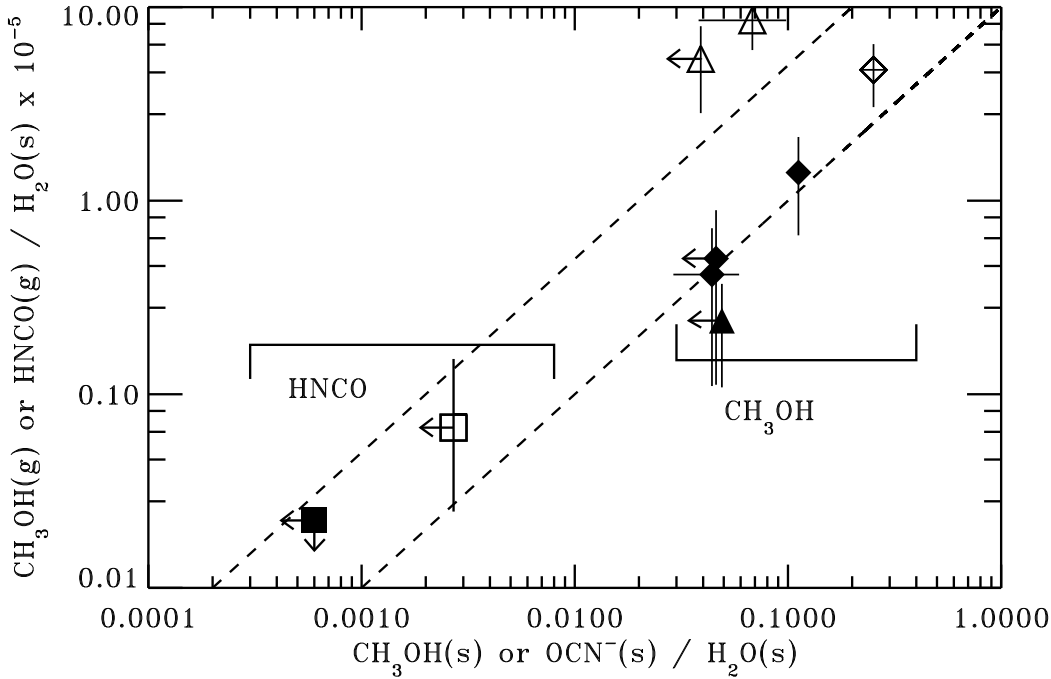}}
\caption{The correlation between ice and gas phase abundances of CH$_3$OH (diamonds and triangles) and HNCO (squares). The filled symbols represent quiescent gas ($<$1.3 km s$^{-1}$) and the open symbols non-quiescent gas ($>$2.4 km s$^{-1}$). In the case of CH$_3$OH, the diamonds represent measurements from this study and triangles from the literature.  The dashed lines show constant CH$_3$OH(gas)/CH$_3$OH(solid) or HNCO(gas)/OCN$^-$(solid) ratios of $1\times10^{-4}$ and $5\times10^{-4}$.}
\label{corr_ch3oh}
\end{figure}

Figure~\ref{corr_ch3oh} shows the correlation between gas and ice abundances of CH$_3$OH and HNCO, including the literature sources. The abundances are with respect to the H$_2$O ice column density, which was found to correlate well with the cold dust column density \citep{Whittet01}. Hence this is a reasonable normalization factor for the lines of sight with quiescent gas, here defined to be line widths $\lesssim1$ km s$^{-1}$, that originates in the cold envelope. It is not a priori a good normalizer for sources with non-quiescent emission, but for consistency the same normalization method is used for all sources. Figure \ref{corr_ch3oh} illustrates a possible correlation between the gas and ice abundances and upper limits for the quiescent sources -- the correlation is not statistically significant due to the many upper limits. The two open triangles in the figure are the RCrA sources, whose higher ratio of gas to ice phase abundance can be attributed to an enhanced radiation field in the region \citep{vanKempen08}. The measured average gas to solid abundance ratio is $1.2\times10^{-4}$ for the quiescent gas. This probably underestimates the true gas to ice ratio because the absorption and emission observations differ, i.e., the gas phase observations probe on average less dense regions than the ice observations.  

\begin{table}
\begin{center}
\caption{The calculated temperatures and column densities.}             
\label{gasdata}      
\centering                          
\begin{tabular}{l c c c c }        
\hline\hline                 
Source&Molecule&Line width&$T_{\rm rot}$&$N_{\rm X}$\\    
&&(km s$^{-1}$)&(K)&($\times10^{13}$ cm$^{-2}$)\\
\hline      
IRAS 03254&CH$_3$OH&0.42--0.49&6$\pm$2&1.8$\pm$1.8\\
&HNCO&0.77&$\sim$4&$\sim$0.48\\
B1-b&CH$_3$OH&0.77--0.88&7$\pm$1&25$\pm$13\\
&HNCO&0.86&$\sim$7&$\sim$2.4\\
L1489 IRS&CH$_3$OH&0.75-0.97&6$\pm$2&1.8$\pm$1.3\\
&HNCO&--&$\sim$6&$<$0.095\\
SVS 4-5&CH$_3$OH&2.4--4.0&7$\pm$1&27$\pm$10\\
&HNCO&2.6--2.9&9$\pm$6&0.38$\pm$0.38\\
\hline
Elias 29$^{\rm a}$&CH$_3$OH&1.3&$\sim9$&$\sim$0.73\\
R CrA 7A$^{\rm b}$&CH$_3$OH&2.4--3.0&18$\pm$2&59$\pm$28\\
R CrA 7B$^{\rm b}$&CH$_3$OH&2.1--2.6&19$\pm$1&94$\pm$28\\
\hline                                   
\end{tabular}
\end{center}
$^{\rm a}$\citet{Buckle02}
$^{\rm b}$derived from \citet{Schoier06}  using rotational diagrams.
\end{table}

\section{Discussion}

Ice photodesorption predicts gas to ice ratios of $10^{-4}-10^{-3}$ (see Appendix) for typical photodesorption yields and envelope conditions. The measured gas to ice ratio in this study of $1.2\times10^{-4}$ agrees well with this prediction, when accounting for the fact that the measured ratio probably underestimates the true gas to ice ratio. The dashed lines in Fig. \ref{corr_ch3oh} further show that all quiescent detections and upper limits are consistent with gas to ice ratios of $(1-5)\times 10^{-4}$. This agreement and the narrowness of the line widths, supports the interpretation that the emission in these lines of sight originates in the cold, quiescent envelope. It also demonstrates that photodesorption alone is sufficient to release the observed ice into the gas, even though other non-thermal processes are not excluded since lack of experimental studies on most non-thermal desorption pathways prevents us from quantifying their relative importance. Gas phase reactions can however be excluded since there is no efficient gas phase reaction pathway to form either CH$_3$OH or HNCO at the observed abundances \citep[][Hassel private comm.]{Garrod07}.

The tentative correlation between gas and ice phase abundances in this pilot study supports the idea that it is possible to determine ice composition by observing the small fraction of the ice that is non-thermally released into the gas phase. To show conclusively that this method works, the size of the sample studied here must be increased and the uncertainty in the derived column densities must be reduced by observing more emission lines. It is also important to remember that until photodesorption data are available for all potential ice species there will be at least a factor of two uncertainty in ice composition estimates using this method due to the different break-up probabilities of different molecules during photodesorption \citep{Oberg08b, Oberg09}. Nonetheless, the method presented here represents a significant improvement on the current lack of observational tools to study complex ices in quiescent regions.

\begin{acknowledgements}
We thank Ruud Visser for stimulating discussions. Funding is provided by NOVA, the Netherlands Research School for Astronomy, the European Early Stage Training Network (`EARA' MEST-CT-2004-504604), a Netherlands Organisation for Scientific Research (NWO) Spinoza grant and the European Community's sixth Framework Programme under RadioNet (R113CT 2003 5058187).
\end{acknowledgements}

\bibliographystyle{aa}

\Online

\begin{appendix} 
\section{Derivation of gas to ice ratios}

The gas to ice ratio for a particular species in a protostellar envelope can be estimated by assuming a steady-state between photodesorption and freeze-out:

\begin{eqnarray}
Y_{\rm pd}\times I_{\rm UV} \times \sigma_{\rm gr}\times f_{\rm x} = 4.57\times 10^4 \times \left(\frac{T}{m_{\rm x}}\right)^{\frac{1}{2}}\times \sigma_{\rm gr} \times n^g_{\rm x}\\
f_{\rm x} = \frac{n^{\rm i}_{\rm x}}{n^{\rm i}}
\end{eqnarray}

\noindent where $Y_{\rm pd}$ is the photodesorption yield set to be $(1-3)\times10^{-3}$ photon$^{-1}$ from our experiments, $I_{\rm UV}$ is the cosmic-ray-induced UV field of $10^4$ photons cm$^{-2}$ s$^{-1}$ and $\sigma_{\rm gr}$ is the grain cross section. The cosmic-ray-induced UV flux assumes a cosmic ray ionization rate of $1.3\times10^{-17}$ s$^{-1}$. Because photodesorption is a surface process, the photodesorption rate of species x depends on the fractional ice abundance $f_{\rm x}$, which is defined to be the ratio of the number density of species x in the ice, $n^{\rm i}_{\rm x}$, to the total ice number density, $n^{\rm i}$. The freeze-out rate of species x depends on the gas temperature $T$, which is set to 15~K, the molecular weight $m_{\rm x}$, and the gas number density $n^{\rm g}_{\rm x}$. For an average molecular weight of 32, this results in a gas phase abundance $n^{\rm g}_{\rm x}/n_{\rm H}$ of $(3-9)\times 10^{-4}f_{\rm x}/n_{\rm H}$.  From this and an average total ice abundance $n^{\rm i}/n_{\rm H}$ of $10^{-4}$, the predicted gas to ice phase abundance ratio is:

\begin{equation}
\frac{n^{\rm g}_{\rm x}}{n^{\rm i}_{\rm x}}\sim \frac{(3-9)\times 10^{-4}/n_{\rm H}\times f_{\rm x}}{n^{\rm i}/n_{\rm H}\times f_{\rm x}}\sim (3-9)/n_{\rm H}
\end{equation}

For a typical envelope density of 10$^4$ cm$^{-3}$, ice photodesorption hence predicts a gas to ice ratio of $10^{-4}-10^{-3}$. The derivation of a gas to ice ratio from observed cold gas emission
lines and ice absorption features in the same line of sight is complicated by the fact that different regions can contribute by varying amounts.  The emission features trace gas in the envelope and cloud both in front and behind the protostar, while the ice absorption features only trace envelope material directly in front of the protostar. The column is hence twice as long for the gas observations. This is probably more than compensated for by using beam averaged gas column densities, as is done in this study, because the large beam traces on average less dense material compared to the pencil beam of the ice absorption observations. Note also that the CH$_3$OH ice abundance may vary between lower and higher density regions. To quantify a conversion factor between the observed and true gas to ice ratio requires detailed modeling of each source which is outside the scope of this study. Instead we here assume that the observed ratio is a lower limit to the true ratio.
 
\end{appendix}

\end{document}